\documentclass[12pt]{article}

\usepackage{amsmath,amssymb}
\usepackage{slashed}
\usepackage{xcolor} 
\usepackage{graphicx}
\usepackage{url}
\usepackage{cite}

\usepackage{verbatim}
   \allowdisplaybreaks
\makeatletter
\@addtoreset{equation}{section}
\makeatother

\graphicspath{{./Figures/}}

\newcommand{\be}{\begin{equation}} 
\newcommand{\ee}{\end{equation}} 
\newcommand{\bea}{\begin{eqnarray}}  
\newcommand{\eea}{\end{eqnarray}}
\newcommand{\bs}{\begin{split}} 
\newcommand{\es}{\end{split}}

\begin{document}
\thispagestyle{empty}

\begin{center}
\hfill  
\begin{center}

\vspace{.5cm}

{\Large\sc Sequestered gravity in gauge mediation}

\end{center}
\vspace{1.cm}

\textbf{ Ignatios Antoniadis$^{\,a,b,c}$, Karim Benakli$^{\,a,b}$,
and Mariano Quiros$^{\,d,e}$}\\

\vspace{.1cm}
${}^a\!\!$ {\em Sorbonne Universit\'es, UPMC Univ Paris 06, UMR 7589, LPTHE, F-75005, Paris, France \\
}

\vspace{.1cm}
${}^b\!\!$ {\em CNRS, UMR 7589, LPTHE, F-75005, Paris, France \\}

\vspace{.1cm}
${}^c\!\!$ {\em Albert Einstein Center, Institute for Theoretical Physics, Bern University\\ Sidlestrasse 5, CH-3012 Bern, Switzerland}

\vspace{.1cm}
${}^d\!\!$ {\em {Institut de Fisica d'Altes Energies (IFAE)\\ The Barcelona Institute of  Science and Technology (BIST)\\ Campus UAB, 08193 Bellaterra (Barcelona) Spain\\ and\\
Instituci\'o Catalana de Recerca i Estudis  
Avan\c{c}ats (ICREA), \\ Campus UAB, 08193 Bellaterra (Barcelona) Spain}}

\vspace{.1cm}
${}^e\!\!$ {\em ICTP South American Institute for Fundamental Research \& Instituto de F\'isica Te\'orica\\ Universidade Estadual Paulista, S\~ao Paulo, Brazil}

\end{center}

\vspace{0.8cm}

\centerline{\bf Abstract}
\vspace{2 mm}
\begin{quote}\small
We present a novel mechanism of supersymmetry breaking embeddable in string theory and  simultaneously sharing the main advantages of (sequestered) gravity and gauge mediation. It is driven by a Scherk-Schwarz deformation along a compact extra dimension, transverse to a $D$-brane stack supporting the supersymmetric extension of the Standard Model. This fixes the magnitude of the gravitino mass, together with that of the gauginos of a bulk gauge group, at a scale as high as $10^{10}$ GeV. Supersymmetry breaking is mediated to the observable sector dominantly by gauge interactions using massive messengers transforming non-trivially under the bulk and Standard Model gauge groups and leading to a neutralino LSP as dark matter candidate. The Higgsino mass $\mu$ and soft Higgs-bilinear $B_\mu$-term could be generated at the same order of magnitude as the other soft terms by effective supergravity couplings as in the Giudice-Masiero mechanism.
\end{quote}

\vfill

\newpage
\section{Introduction}
\label{Introduction} 
The gravity mediated supersymmetry breaking scenario with an $\mathcal O(TeV)$ gravitino~\cite{Nilles:1983ge}, that can be realized for instance in the minimal supersymmetric extension of the Standard Model (MSSM) has, apart from solving the hierarchy problem, the phenomenological advantages of providing gauge coupling unification at scales $M_{GUT}\sim 10^{16}$ GeV and a standard dark matter candidate in the presence of unbroken $R$-parity if the lightest supersymmetric particle (LSP) is a neutralino. Gravity mediation has nevertheless two main drawbacks: \textit{i)} Gravitational interactions are not automatically flavor blind and thus they do not guarantee a solution to the supersymmetric flavor problem; \textit{ii)} An $\mathcal O(TeV)$ gravitino decays in a lifetime of about $10^6$ sec, leading to a huge entropy production after the big bang nucleosynthesis (BBN) and spoiling its predictions unless the reheating temperature after inflation is ~$\lesssim 10^{10}$ GeV, which puts an uncomfortable bound on inflationary scenarios. The latter is known as the \textit{cosmological gravitino problem} \cite{Moroi:1993mb}.
 
The main motivation for gauge mediated supersymmetry breaking (see for example \cite{Giudice:1998bp} and references therein) is that it provides flavor independent soft breaking terms thus avoiding strong experimental constraints on flavor changing neutral currents. On the other hand, it also has some problematic drawbacks: \textit{i)} One loses the standard dark matter candidate as a WIMP (weakly interacting massive particle) since the LSP is now the gravitino; \textit{ii)} The gravitino can be (warm) dark matter only if its mass is $m_{3/2}\lesssim 1$ keV, which requires an upper bound on the messenger mass, $M$, over its number, $N$, as $M/N\lesssim 10^7$ GeV, while for larger gravitino masses the reheating temperature after inflation is strongly constrained; \textit{iii)} There is no compelling way to generate a  supersymmetric $\mu$-parameter (Higgsino mass) and a $B_\mu$ soft-term (Higgs bilinear) of the same order as the other soft terms. Although none of these problems can disqualify gauge mediation as a very appealing mechanism of supersymmetry breaking transmission to the observable sector it would be certainly interesting to find a theory where these problems do not appear.

In this work, we propose a mechanism to solve these problems by appropriately sequestering the supersymmetry breaking in the hidden sector in a string theory context. Due to the sequestering of gravity, even if our gravitino is supermassive gauge mediated interactions will be dominant over gravity mediated ones thus solving the supersymmetric flavor problem. Moreover since the gravitino mass is $m_{3/2}\gg 1$ TeV our model does not exhibit any gravitino problem and moreover the best candidate for dark matter is the lightest neutralino.

Our basic setup is the following. We consider the MSSM localized in three (spatial) dimensions, on a collection of $D$-brane stacks that we call in short $D3$-brane, transverse to a ``large" extra dimension on a semi-circle (orbifold) of radius $R$, along which there is a bulk ``hidden" gauge group $G_H$. We assume that $G_H$ has a non-chiral spectrum. There are in general matter fields, described by excitations of open strings stretched between the Standard Model (SM) brane and the hidden group $D$-brane extended in the bulk, and thus localized in their three-dimensional (spatial) intersection. They transform in the corresponding bi-fundamental representations and, since they are non-chiral, they can acquire a mass $M$ by appropriate brane displacements (or equivalently Wilson lines), that we consider as a parameter of the model.  They will play the role of messengers to transmit supersymmetry breaking to the observable sector.

Supersymmetry breaking is induced by a Scherk-Schwarz (SS) deformation along the extra dimension generating a Majorana mass for fermions in the bulk, namely the gravitino and the gauginos of $G_H$, proportional to the compactification scale $1/R$, but leaving the SM brane supersymmetric~\cite{Antoniadis:1998ki}. The breaking is mediated to the observable SM sector by both gravitational~\cite{Antoniadis:1997ic,Antoniadis:1998ki} and gauge interactions~\cite{Benakli:1998pw} (via the bi-fundamental messenger fields), whose relative strength is controlled by the compactification scale and messenger mass. Fixing for definiteness the MSSM soft terms at the TeV scale and requiring the gravitational contribution to be suppressed with respect to the gauge mediated one by at least one order of magnitude~\footnote{Notice that as pure SS gravity mediation is flavor diagonal, one order of magnitude suppression in the scalar masses is enough to make gauge mediation the dominant mechanism without conflicting with any supersymmetric flavor problem.}, one finds that the compactification scale $1/R$ should be less than about $1/R \sim 10^{10}$~GeV, corresponding to a string scale in the unification region $M_{GUT}\sim 10^{16}$~GeV, inferred by extrapolating the low energy SM gauge couplings with supersymmetry. 

The resulting MSSM soft terms (sfermion and gaugino masses) have then the usual pattern of gauge mediation, with in particular scalar masses that are essentially flavor blind, guaranteeing the absence of dangerous flavor changing neutral current interactions. Note thought that in contrast to the standard gauge mediation scenario, the gravitino mass is heavy, of order the compactification scale, evading the gravitino overproduction problem and having as LSP the lightest neutralino, like in models of gravity mediation, in the right ballpark needed for describing the missing dark matter of the Universe required by astrophysical and cosmological observations. On the other hand, a globally supersymmetric Higssino mass $\mu$-term and its associated Higgs-bilinear soft term $B_\mu$ can be generated in a similar way as in the Giudice-Masiero mechanism~\cite{Giudice:1988yz}, by effective supergravity $D$-term interactions involving a non-holomorphic function depending on the radius modulus field $T$ whose $F$-auxiliary component acquires a non-vanishing expectation value, as dictated by the SS deformation. Under a reasonable assumption on the asymptotic dependence, the induced $\mu$ and $B_\mu$ parameters are of the same order with the rest of the MSSM soft supersymmetry breaking terms.

The outline of the paper is the following. In Sec.~2 we compute the contribution of gravity mediation to scalar and gaugino masses and we define the upper bound of the compactification scale in order to suppress the former contribution from the total. In Sec.~3 we compute the corresponding contribution of gauge mediation and determine the region of messenger mass that leads to a viable phenomenological spectrum. In Sec.~4 we discuss the generation of $\mu$ and $B_\mu$ terms, while Sec.~5 contains our conclusions. Finally, in App.~A, we present the details of the computation of the induced $F$-auxiliary expectation value in the messenger sector off-shell, needed for the evaluation of the gauge mediated contributions in the main text.

\section{Gravity mediation for SS supersymmetry breaking}
\label{gravitational}
Our starting setup is a higher dimensional space where the MSSM is localized on a D3-brane that is perpendicular to a large compact coordinate (of radius $R$). Supersymmetry is assumed to be broken by a Scherk-Schwarz~\cite{Scherk:1978ta} mechanism giving to the gaugino $\lambda_H$ and gravitino Kaluza-Klein (KK) modes a common mass:
\be
M_{n}(\omega)=m_{3/2}+\frac{n}{R},\quad m_{3/2}=\frac{\omega}{R}
\label{gravmasses}
\ee
where $m_{3/2}$ is the mass of the gravitino zero mode and $\omega$ a real parameter $0<\omega<\frac{1}{2}$. 

We are interested here in evaluating the size of the MSSM supersymmetry breaking soft terms mediated by gravitational (grav) effects $(m_0^{\rm grav },M_{1/2}^{\rm grav})$ that we shall compare in the next section to those from gauge interactions. Supersymmetry breaking is transmitted from the bulk to the brane by one-loop gravitational interactions giving a (squared) mass to scalars proportional to~\cite{Antoniadis:1997ic,Gherghetta:2001sa}
\be
\left(m_0^{\rm grav}\right)^2=\frac{1}{M_P^2}\sum_n\int \frac{d^4k}{(2\pi)^4}k^2\left[\frac{1}{k^2+M_n^2(\omega)}-\frac{1}{k^2+M_n^2(0)}\right]
\label{m0grav}
\ee
where $M_P=2.4\times 10^{18}$ GeV is the reduced Planck mass, and a Majorana mass to gauginos proportional to
\be
M_{1/2}^{\rm grav}=\frac{1}{M_P^2}\sum_n\int \frac{d^4k}{(2\pi)^4}\left[\frac{M_n(\omega)}{k^2+M_n^2(\omega)}-\frac{M_n(0)}{k^2+M_n^2(0)}\right]
\label{m12grav}
\ee

The gravitational squared mass of scalars can then be expressed as
\be
\left(m_0^{\rm grav}\right)^2=-\frac{1}{R^2}\frac{1}{(M_PR)^2}\frac{1}{(4\pi)^2} f_0(\omega)
\ee
where
\be
f_0(\omega)=\frac{3}{2 \pi^4}\left[2\zeta(5)-Li_5(e^{2i\pi\omega})-Li_5(e^{-2i\pi\omega}) \right]
\ee
and the gravitational Majorana gaugino mass as
\be
M_{1/2}^{\rm grav}=\frac{1}{R}\frac{1}{(M_PR)^2}\frac{1}{(4\pi)^2} f_{1/2}(\omega)
\ee
where
\be
f_{1/2}(\omega)=\frac{3i}{8\pi^3}\left[Li_4(e^{-2i\pi\omega})-Li_4(e^{2i\pi\omega})  \right]
\ee
\begin{figure}[htb]
 \begin{tabular}{cc}
 \includegraphics[width=8.1cm]{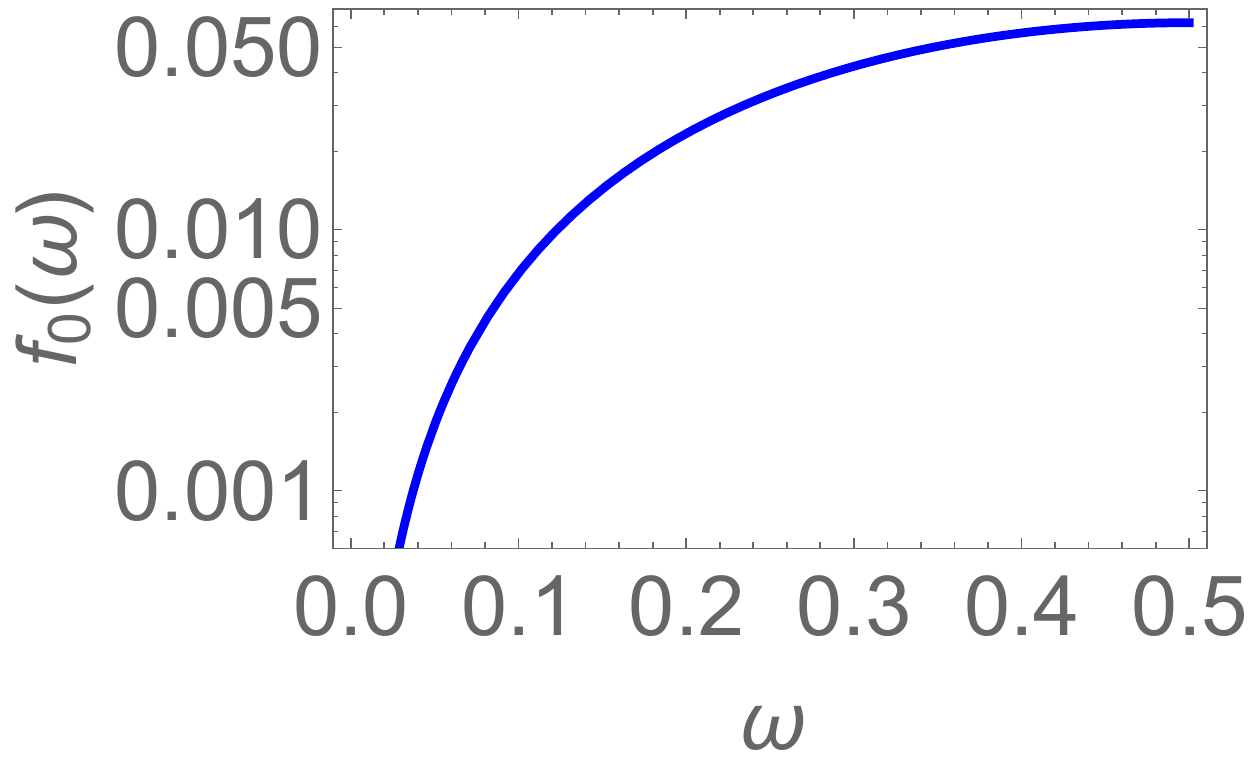}  
  \includegraphics[width=8.1cm]{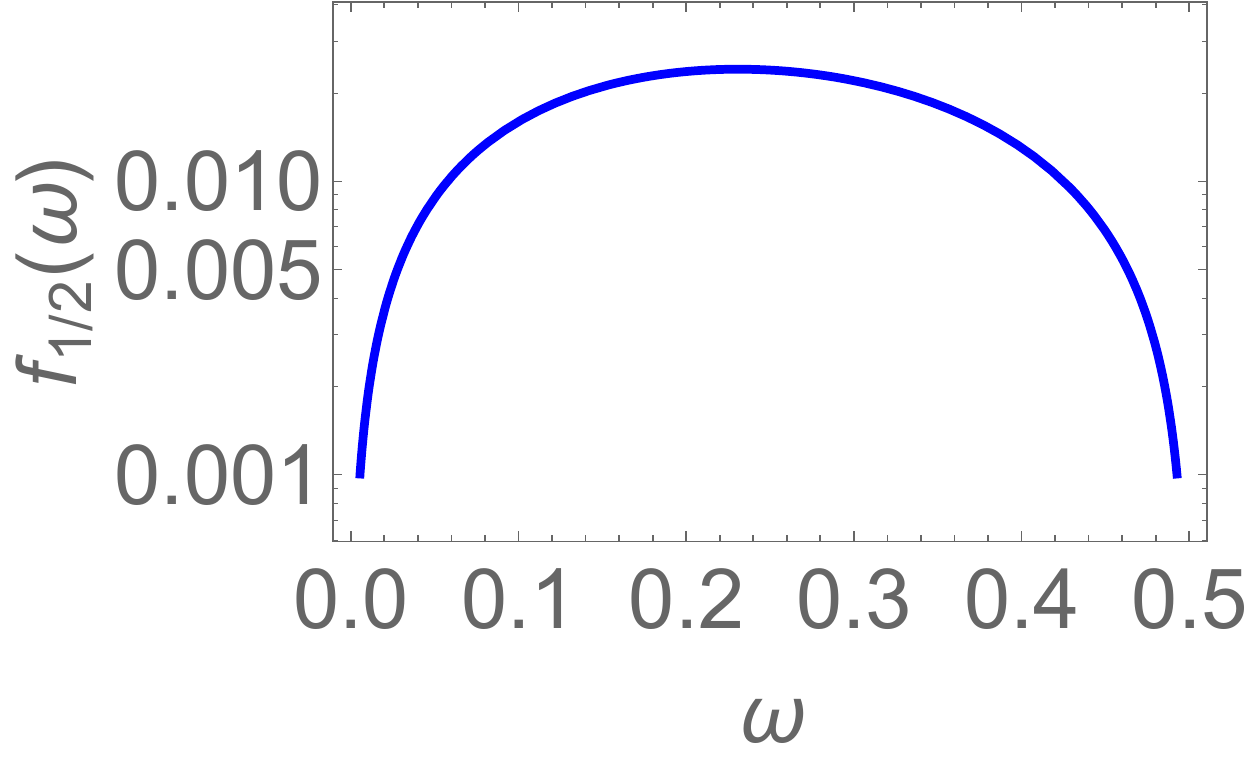}  
\end{tabular}
\caption{\it Left panel: Plot of $f_0(\omega)$. Right panel: Plot of $f_{1/2}(\omega)$}
\label{fig1}
\end{figure}
Note that $f_{1/2}(\omega)$ vanishes for $\omega = \frac{1}{2}$ because of a preserved R-symmetry~\cite{Antoniadis:1997ic,Gherghetta:2001sa}. As we focus on models with Majorana masses for gauginos, we need therefore to take $\omega \in  ] 0, 1/2 [ $. For $\omega \sim 1/2$ there are two quasi-degenerate Majorana gravitinos, one of which couples to the MSSM (the even parity one) at the three-brane location.  The functions $f_0(\omega)$ and $f_{1/2}(\omega)$ are plotted in Fig.~\ref{fig1}, left and right panels respectively, where we can see that their values are typically $\mathcal O(10^{-2})$. 

\begin{figure}[htb]
\begin{center}
 \includegraphics[width=10cm]{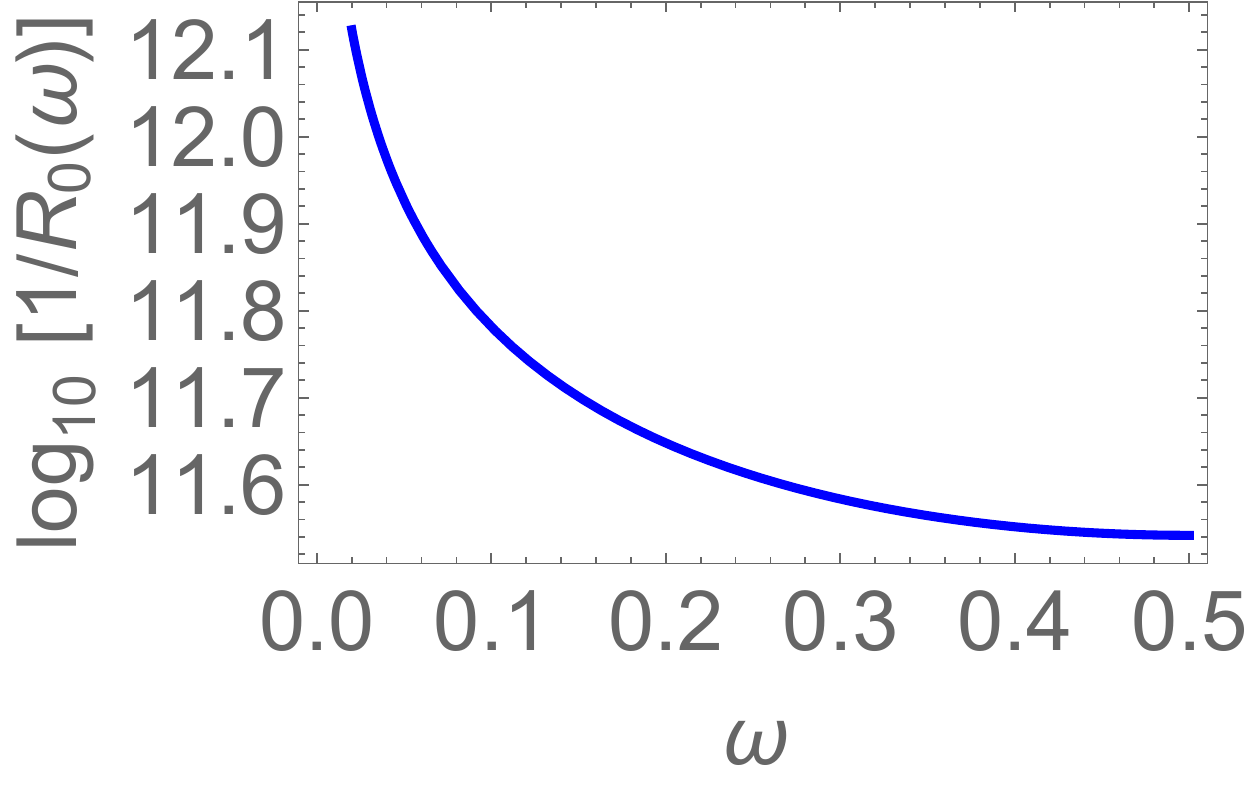}  
 \end{center}
\caption{\it Plot of $\log_{10}[1/R_0(\omega)/{\rm GeV}]$ by fixing $|m_0^{\rm grav}|=1$ TeV.}
\label{fig2}
\end{figure}

The gravitational contributions to the scalar squared masses are (in the absence of other moduli mediating supersymmetry breaking) negative and flavor diagonal (assuming a Kahler matter metric flavor diagonal) as they are mediated by diagrams with gravitinos and gravitons exchanged in the loops~\cite{Antoniadis:1997ic}.  
One can always fix the radius to a reference value $R_0$ by imposing the condition that $|m_0^{\rm grav}|\simeq 1$ TeV. The result $1/R_0(\omega)$ is plotted in Fig.~\ref{fig2} where we can see that $(1/R_0)\sim 10^{12}$~GeV (almost) independently of the value of $\omega$.  For the corresponding values of the gravitino mass [$m_{3/2}\equiv \omega /R_0(\omega)$] the Majorana gaugino masses are small enough to be neglected. For larger radii the gravitational contributions to scalar masses can become negligible, so if there exists another (gauge) mediation mechanism of supersymmetry breaking to the observable sector which generates leading contributions to scalar masses, as we will describe in Sec.~\ref{gauge}, a one order of magnitude suppression of the gravitational contribution with respect to the leading one should be enough~\footnote{In this paper we are assuming flavor diagonal matter metric. For more general matter metrics (non-diagonal in flavor space) a stronger suppression (around two orders of magnitude) should be required and our results should be slightly modified as we will see.}. 

There is also a contribution that originates from anomaly mediation (AM) and that provides scalar and gaugino masses as~\cite{Randall:1998uk,Giudice:1998xp}
\be
m_0^{AM}\sim M_{1/2}^{AM}\sim \frac{\lambda^2}{16\pi^2}F_\phi
\label{AM}
\ee
where $\lambda$ indicates the different four-dimensional gauge and Yukawa couplings and $F_\phi$ is the $F$-component of the chiral compensator $\phi=1+\theta^2 F_\phi$. In the case of a Scherk-Schwarz breaking it turns out that, as in no-scale models, $F_\phi\ll m_{3/2}$. In fact one can prove that at the tree level $F_\phi=0$~\cite{Marti:2001iw} while at one-loop a gravitational Casimir energy $V\sim (1/16\pi^2) 1/R^4$ is generated which yields a non-zero value of $F_\phi$ as $F_\phi\sim (1/16\pi^2) m_{3/2}^3/M_P^2$~\cite{ArkaniHamed:2004yi}. Therefore its contribution to scalar masses from (\ref{AM}) is negligible, as compared to the gravitational contribution, Eq.~(\ref{m0grav}), while its contribution to gaugino masses is of the same order of magnitude as the gravitational ones, Eq.~(\ref{m12grav}).

\section{Gauge mediation for SS supersymmetry breaking}
\label{gauge}
We now turn to compute the supersymmetry breaking effects induced by the bulk-brane gauge interactions. We will denote as $G_H$ the gauge symmetry group living in the bulk and by $\alpha_H$ the associated four-dimensional coupling~\footnote{$\alpha_H$ is the dimensionless four-dimensional coupling of gauge fields on a $D4$-brane wrapped along a cycle of length $\pi R$ in the extra dimension.}.
We  consider a number of messengers $(\phi_I,\bar\phi_I)$ living in the intersection between the MSSM $D3$-brane and the brane in the bulk which contains the SS direction. The messengers transform under the representation $(\mathcal R^{G_H}_{\phi_I},\, \overline{\mathcal R}^{G_H}_{\phi_I})$ of the hidden gauge group $G_H$  and under the representation $(\mathcal R^{G_{SM}}_{\phi_I},\, \overline{\mathcal R}^{G_{SM}}_{\phi_I})$ of the Standard Model group $G_{SM}$. 

\subsection{The messenger sector}

By the SS supersymmetry breaking the gauginos of the hidden gauge group $\lambda_H$ acquire, as the gravitinos,  soft Majorana masses: 
\be
M_{n}(\omega)=M_{1/2}+\frac{n}{R},\quad M_{1/2}=m_{3/2}=\frac{\omega}{R}
\label{Hgauginomasses}
\ee

We assume, as in ordinary gauge mediation, that the messengers have a supersymmetric mass as
\be
W=\phi_I\,M_{IJ}\,\bar\phi_J
\ee
where, without loss of generality, the mass matrix is diagonal $M_{IJ}=M_I\delta_{IJ}$, and we are assuming that $M_I\simeq M$ ($\forall I$). This diagonal mass matrix will induce a diagonal supersymmetry breaking parameter, denoted by $F_I=\lambda_I F$, through the one-loop radiative corrections induced by the bulk gaugino $\lambda_H$ and Dirac fermion $(\widetilde \phi_I,\widetilde{\bar\phi}_I)$, that is proportional to~\cite{Delgado:2001si}~\footnote{In fact the value of the parameter $F$ actually depends on the value of the external momentum $q$, which we are taking here at $q=0$. The complete, more technical, analysis where the momentum-dependent supersymmetry breaking parameter $F(q)$ is used to compute scalar and gaugino masses is left for Appendix~A where we will demonstrate that the approximation of considering the insertion parameter $F(0)$ is good provided that the mild condition $M\lesssim 0.1/R$ is fulfilled.}
\be
F=\frac{\alpha_H}{\pi} \, \int_0^\infty dp^2\, p^2\sum_n\left[\frac{M_n(\omega)}{p^2+M_n^2(\omega)} -\frac{M_n(0)}{p^2+M_n^2(0)} \right] \frac{M}{p^2+M^2}
\label{integral}
\ee
where $\lambda_I=C^{G_H}_{\phi_I}$ is the quadratic Casimir of the representation $\mathcal R^{G_H}_{\phi_I}$ of $G_H$. The integral has different behaviors for $RM<1$ and $RM>1$. In fact it can be written as
\be
F=\frac{\alpha_H}{4\pi} \frac{M}{R}\left\{
\begin{tabular}{lcl}
$g_0(\omega)$, & for $RM\ll 1$:& $g_0(\omega)=2i\left[Li_2(r)-Li_2(1/r) \right]/\pi$ \\
&&\\
$1/\left(MR \right)^2\, g_\infty(\omega)$, & for $RM\gg 1$:& $g_\infty(\omega)=3i\left[Li_4(r)-Li_4(1/r) \right]/\pi^3$
\end{tabular}
\label{F}
\right.
\ee
where $r=e^{-2i\pi\omega}$.
\begin{figure}[htb]
 \begin{tabular}{cc}
 \includegraphics[width=7.8cm]{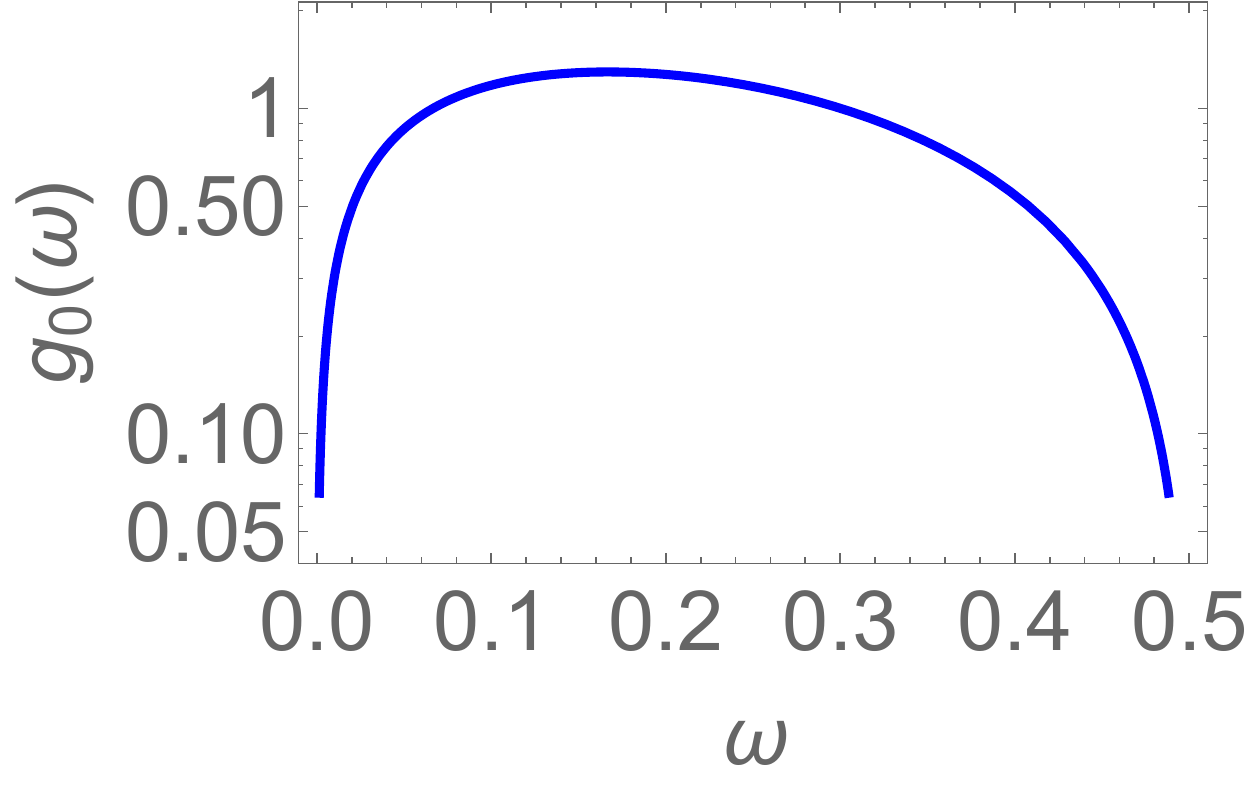}  
  \includegraphics[width=7.8cm]{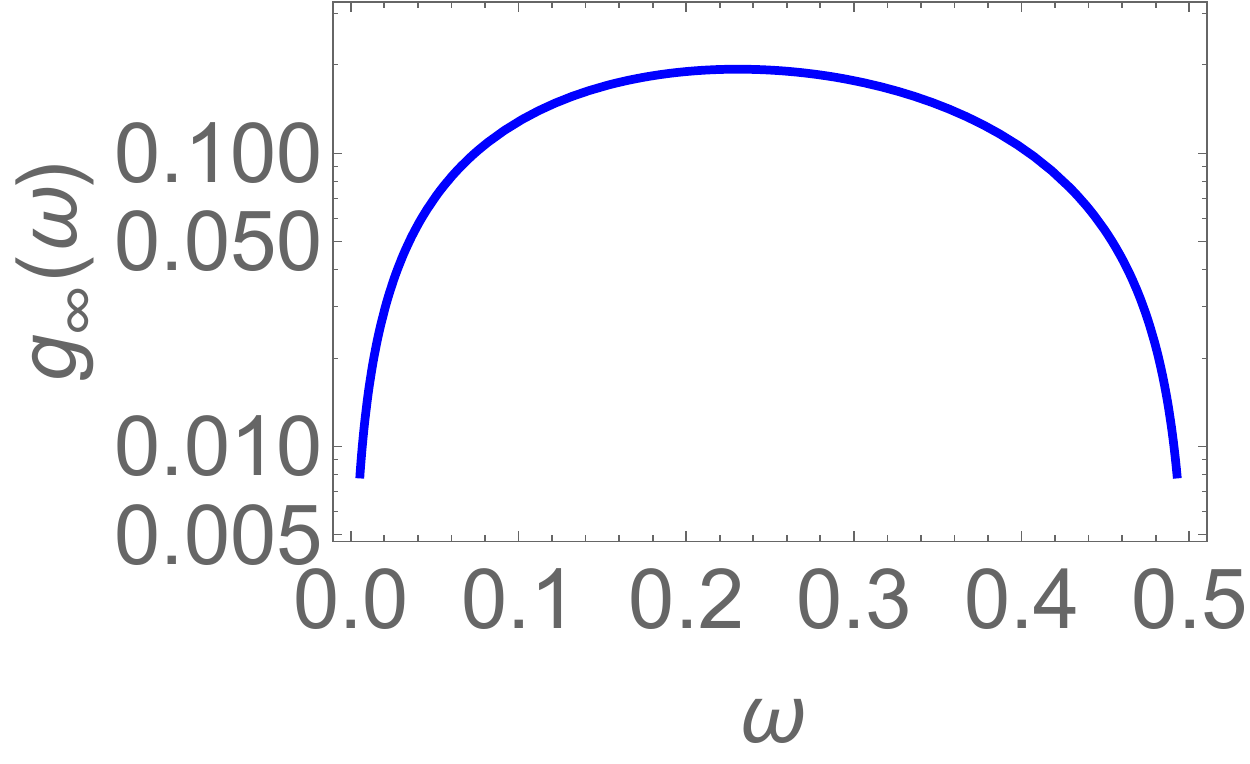}  
\end{tabular}
\caption{\it Left panel: Plot of $g_0(\omega)$. Right panel: Plot of $g_{\infty}(\omega)$}
\label{fig3}
\end{figure}
The functions $g_0(\omega)$ and $g_\infty(\omega)$ are plotted in Fig.~\ref{fig3} where we see that they satisfy the relation $g_\infty(\omega)\sim 0.1 g_0(\omega)$. A quick glance at Eq.~(\ref{F}) shows that the parameter $F/M$ is larger for $M\lesssim1/R$ than for $M\gtrsim 1/R$ so that in the following we will only consider the former case of $g_0(\omega)$. 

\subsection{The observable sector}

Supersymmetry breaking is then transmited through the usual gauge mediation mechanism~\cite{Giudice:1998bp} to squark, slepton and gaugino masses of the MSSM. In order to not spoil the MSSM gauge coupling unification we can assume that the messenger sector consists in complete $SU(5)$ representations. For instance we can assume $n_5$ multiplets in the $5+\bar 5$, $[(D,L)+(\bar D,\bar L)]$, and $n_{10}$ multiplets in the $10+\overline{10}$, $[(Q,U,E)+(\bar Q,\bar U,\bar E)]$. The mass generated by gauge mediation for gauginos and scalars can then be written as
\begin{align}
M_3&=\frac{\alpha_3}{4\pi}\left[ n_5\Lambda_D+n_{10}(2\Lambda_Q+\Lambda_U )  \right]\nonumber\\
M_2&=\frac{\alpha_2}{4\pi}\left[ n_5\Lambda_L+3n_{10}\Lambda_Q  \right]\nonumber\\
M_1&=\frac{\alpha_1}{4\pi} \frac{6}{5}\left[ n_5\left(\frac{1}{3}\Lambda_D +\frac{1}{2}\Lambda_L \right)+n_{10}\left(\frac{1}{6}\Lambda_Q+\frac{4}{3}\Lambda_U+\Lambda_E  \right)\right]
\end{align}
where $\Lambda_I= C_I^{G_H} F/M$ ($I=D,L,Q,U,E$), and
\be
m_{\tilde f}^2=2\left[C_3^{\tilde f}\left(\frac{\alpha_3}{4\pi}\right)^2 \Lambda_3^2+C_2^{\tilde f}\left(\frac{\alpha_2}{4\pi}\right)^2 \Lambda_2^2+C_1^{\tilde f}\left(\frac{\alpha_1}{4\pi}\right)^2 \Lambda_1^2
\right]
\ee
where $C_i^{\tilde f}$ is the quadratic Casimir of $\tilde f$ representation with the normalization $C_1^f=\frac{3}{5}Y_f^2$, and where all couplings are at the scale $M$. Similarly the scales $\Lambda_i$ are defined as
\begin{align}
\Lambda_3^2&=n_5\Lambda_D^2+n_{10}(2\Lambda_Q^2+\Lambda_U^2 )\nonumber\\
\Lambda_2^2&= n_5\Lambda_L^2+3n_{10}\Lambda_Q^2\nonumber\\
\Lambda_1^2&=\frac{6}{5}\left[ n_5\left(\frac{1}{3}\Lambda_D^2 +\frac{1}{2}\Lambda_L^2 \right)+n_{10}\left(\frac{1}{6}\Lambda_Q^2+\frac{4}{3}\Lambda_U^2+\Lambda_E^2  \right)\right]
\end{align}

To simplify the analysis we will assume that the structure of $SU(5)$ multiplets is not spoiled by $G_H$ so that $\Lambda_I\equiv \Lambda_5=C_5^{G_H}\, F/M$ ($ I=D,L$) and $\Lambda_I\equiv \Lambda_{10}=C_{10}^{G_H}\, F/M$ ($ I=Q,U,E$)~\footnote{This amounts to assuming that all messengers in a given representation \textbf{r} of $SU(5)$ are in the same representation of the hidden group $G_H$ with quadratic Casimir $C_{\textbf{r}}^{G_H}$. } in which case the previous equations yield 
\begin{align}
M_i&=\frac{\alpha_i}{4\pi}\left[n_5\Lambda_5+3 n_{10}\Lambda_{10} \right]\nonumber\\
\Lambda_i^2&=n_5\Lambda_5^2+3 n_{10}\Lambda_{10}^2 
\label{escalas}
\end{align}
As we want the LSP, and thus the dark matter component of the universe, to be a well tempered admixture of Bino/Higgsino ($\tilde B/\tilde H$) we need sfermions to be heavier than $M_1$ at the low scale, a condition which prevents a large number of messengers  (as we will next see). So from Eq.~(\ref{escalas}) it is obvious that the case where only the number $n_5$ of $5+\bar 5$ is charged under $G_H$, i.e.~$C_5^{G_H}\neq 0$, while the messengers in the $10+\overline{10}$ are neutral under $G_H$ and thus $C_{10}^{G_H}=0$ (i.e.~$\Lambda_{10}=0$), is preferred as $n_{10}$ has multiplicity three in (\ref{escalas}). In this case we obtain the usual expressions of minimal gauge mediation
\begin{align}
M_i&=\frac{\alpha_i}{4\pi}\Lambda_G,\quad \Lambda_G=n_5\Lambda_5,\nonumber\\
\Lambda_i^2&=\Lambda_S^2,\quad \Lambda_S^2=n_5\Lambda^2_5
\label{Mi}
\end{align}
and we should keep $n_5$ as small as possible. Moreover to keep the multiplicity of the representations $\mathcal R^{G_H}_5$ to the lowest possible values we will assume that $G_H=U(1)_H$.

\begin{figure}[htb]
\begin{center}
 \begin{tabular}{cc}
 \includegraphics[width=7.9cm]{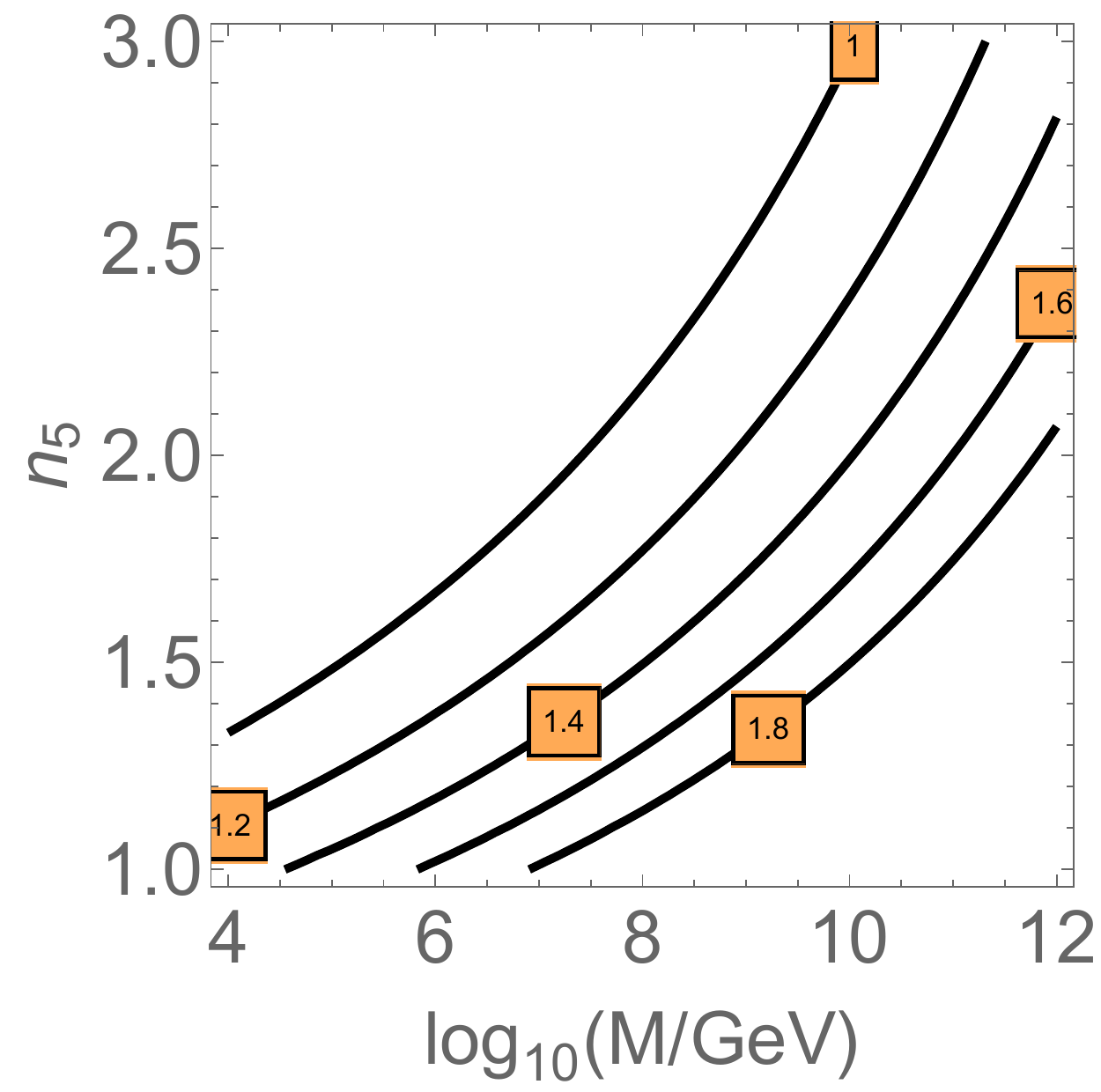}  
  \includegraphics[width=7.5cm]{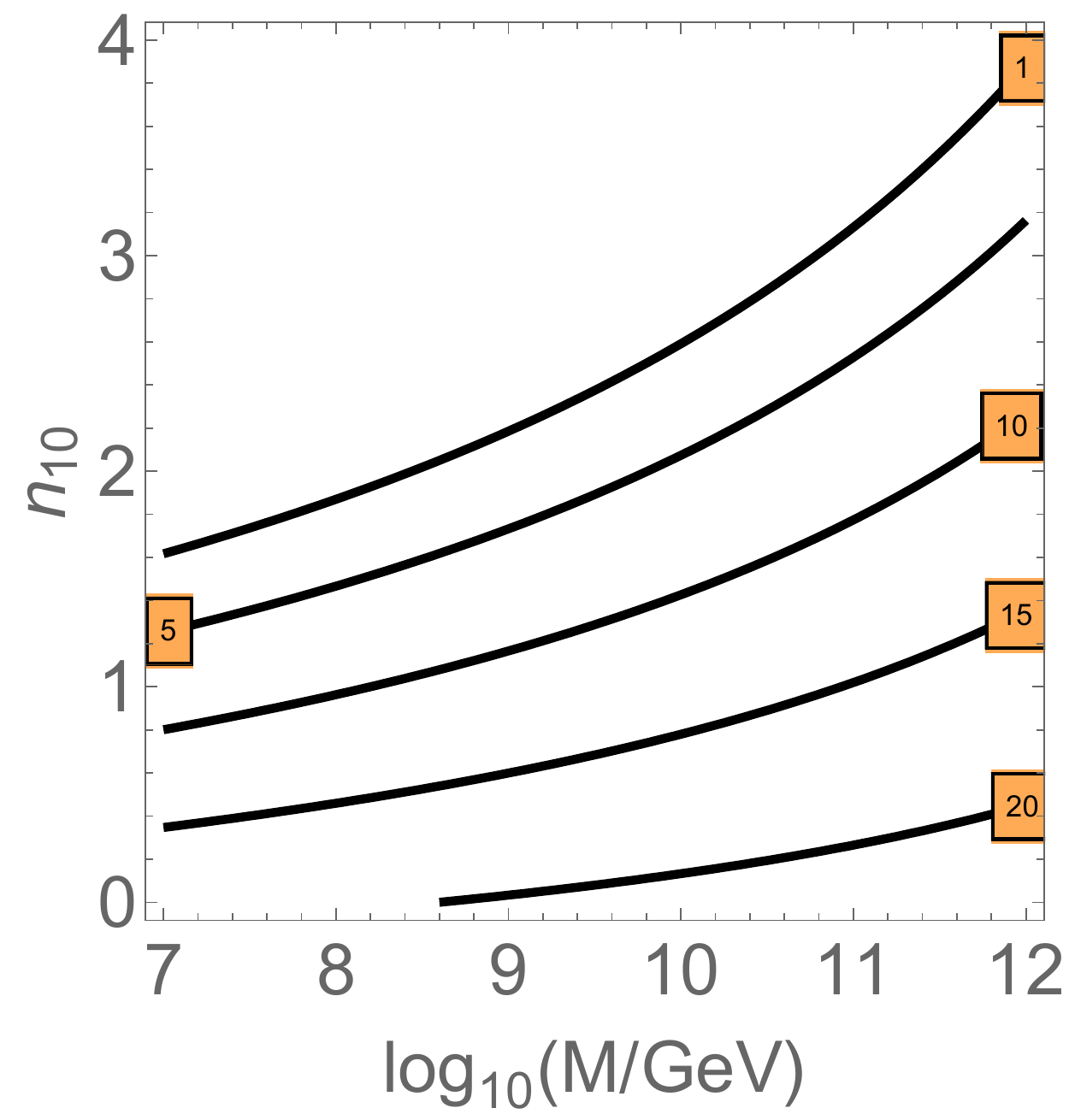}  
\end{tabular}
\caption{\it Left panel: Contour lines of $m^2_{\tilde \ell_R}(\mu_0)/M_1^2(\mu_0)$ in the plane $(\log_{10}M/GeV,n_5)$. Right panel: Contour lines of $1/\alpha_{GUT}$ in the plane $(\log_{10}M/GeV,n_{10})$ for $n_5=2$, where $n_{10}$ is the number of messengers in the $10+\overline{10}$ of $SU(5)$ uncharged under $G_H$.
}
\label{fig4}
\end{center}
\end{figure}

In this framework, as the masses are generated at the scale $M$ and run to the scale $\mu_0\sim \mathcal O(\textrm{TeV})$ by the renormalization group equations, the lightest gaugino is the $U(1)$ Bino and the lightest scalar is the right-handed slepton $\tilde\ell_R$~\footnote{In particular, considering the small effect of the $\tau$ Yukawa coupling, the lightest slepton would be the $\tilde\tau_R$.}. As the lightest supersymmetric particle (LSP) is stable~\footnote{We are here imposing $R$-parity.} we need the Bino ($\tilde B$ with mass $\sim M_1(\mu_0)$) to be lighter than $\tilde\ell_R$ as we already mentioned. In this case we would also need the neutral Higgsino $\tilde H$ Dirac mass $\mu$ (see next section) to be $\mu\sim M_1$ to avoid the over-closure of the Universe and predict the thermal dark matter density measured by WMAP. This is the so-called well tempered $\tilde B/\tilde H$ scenario~\cite{ArkaniHamed:2006mb}. The conditions for the Bino to be lighter than $\tilde \ell_R$ are shown in the left panel of Fig.~\ref{fig4} where we plot contour lines of the ratio $m^2_{\tilde \ell_R}(\mu_0)/M_1^2(\mu_0)$ in the plane ($\log_{10}M/GeV,n_5)$.
We can see that for $M\lesssim 10^8$ GeV there is the bound $n_5\leq 2$. In particular for $n_5=1$ the value of $M$ is not constrained by the LSP requirement, and for $n_5=2$ we have $M\gtrsim 10^7$ GeV. Finally for $n_5=3$ we have $M\gtrsim 10^{10}$ GeV, which essentially saturates the bound obtained in App.~\ref{apendice}, see Eq.~(\ref{cota})~\footnote{We can relax the allowed window by lifting the condition that $G_H$ does not spoil the structure of $SU(5)$ multiplets. A possibility we will not pursue along this paper.}. All these predictions are independent on the value of $n_{10}$ as we are assuming $C_{10}^{G_H}=0$ and thus the $10+\overline{10}$ consists in a supersymmetric sector. As it is shown in the right panel of Fig.~\ref{fig4}, where we show contour lines of constant $1/\alpha_{GUT}$ in the plane $(\log_{10}M/GeV,n_{10})$ for $n_5=2$, their presence will modify the value of the unification gauge coupling $\alpha_{GUT}$ and the model predictions. 

There is another way of avoiding the constrains provided by the left plot of Fig.~\ref{fig4}. It consists in assuming that the LSP is mainly Higgsino-like with a (heavy) mass $\mu (\mu_0)\sim 1$ TeV~\cite{ArkaniHamed:2006mb} so that it would be possible to have $\mu(\mu_0) \lesssim m_{\tilde \ell_R}(\mu_0)\lesssim M_1(\mu_0)$. However in this case $M_1\gtrsim 1$ TeV which would imply, in the minimal gauge mediation scenario we are assuming in this paper, a very heavy gluino $M_3\gtrsim 5$ TeV, and thus a quite heavy supersymmetric spectrum.

To explicitly compute the value of $\Lambda_S$ we will use the fact that $\alpha_H$ is the four-dimensional gauge coupling of $G_H$, so its value is given by~\cite{Polchinski:1998rr} 
\be
\alpha_H=2\alpha_{GUT}/(RM_s)
\ee
where $\alpha_{GUT}$ is the SM coupling at the unification scale, and $M_s$ the fundamental (string) scale. The value of $\alpha_H$ can be written in terms of the reduced 4D Planck scale $M_P=2.4\times 10^{18}$ GeV, using the relation~\cite{Polchinski:1998rr} 
\be
M_P^2=\frac{M_s^3R}{8\pi\alpha_{GUT}^2} 
\label{Planckrelation}
\ee
We can then write
\be
\Lambda_S=K_H g_0(\omega) \left(1/R\right)^{5/3}\left(1/M_P\right)^{2/3}
\ee
where the pre-factor $K_H$ given by
\be
K_H=\sqrt{n_5}\left(\frac{\alpha_{GUT}}{64\,\pi^4}  \right)^{1/3} C^{G_H}_\phi \simeq 
10^{-1.1}\, \sqrt{\frac{n_5}{2}}\, \alpha_{GUT}^{1/3} C^{G_H}_\phi 
\label{KH}
\ee
contains all the model dependence on the hidden sector.
\begin{figure}[htb]
 \begin{tabular}{cc}
 \includegraphics[width=8.5cm]{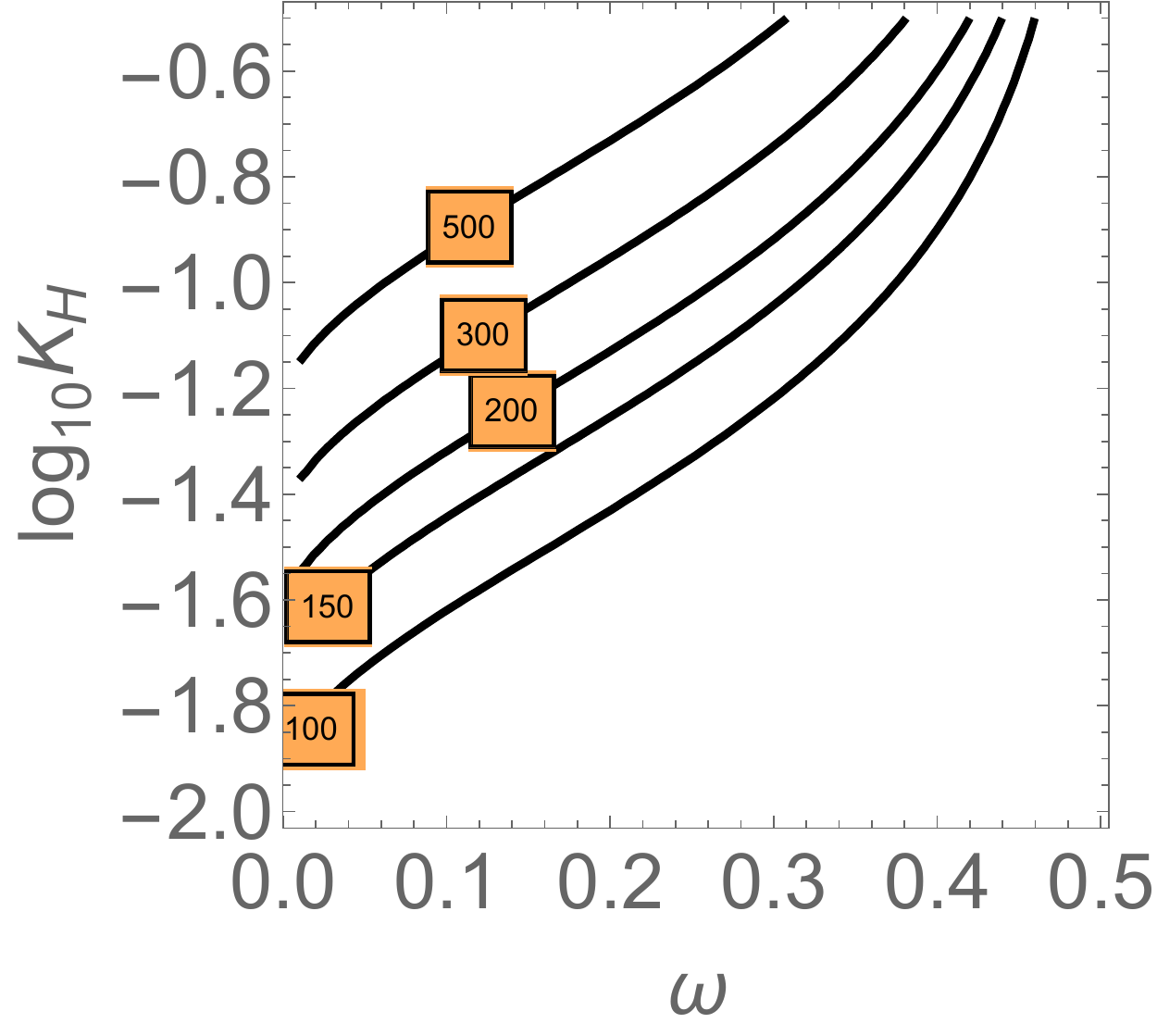}  
 \includegraphics[width=8.5cm,height=7cm]{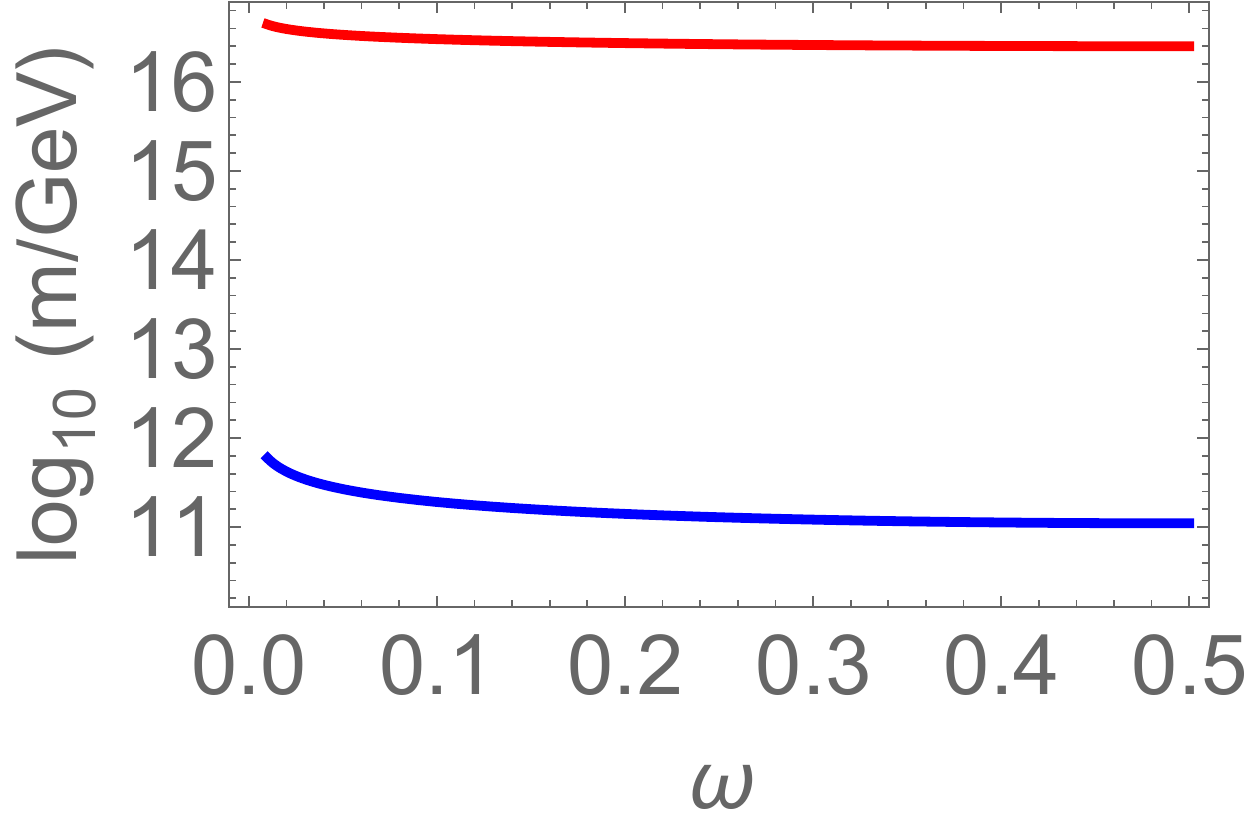}  
\end{tabular}
\caption{\it  Left panel: Contour lines of constant $\Lambda_S$ (in TeV) in the plane $(\omega,K_H)$. Right panel: Plot of $\log_{10}[1/R(\omega)/GeV]$ (lower line) and $\log_{10}[M_s/GeV]$ (upper line) as functions of $\omega$. For the upper line we have considered the case $\alpha_{GUT}=\mathcal O(1)$.}
\label{fig5}
\end{figure}

We will fix the radius as $1/R(\omega)\equiv\sqrt{ \delta} /R_0(\omega)$, where $R_0(\omega)$ is the value of the radius fixed by the plot in Fig.~\ref{fig2} and $\delta$ is the suppression coefficient which should make the gravitational contribution to the scalar masses negligible, as compared to those obtained from gauge mediation. 
In fact for $\delta\lesssim 0.1$ the previous contributions to scalar masses are separated by at least one order of magnitude. Note that at leading order, the soft breaking terms induced by the SS gravitational interactions are flavor blind, thereof a stronger suppression is not required as we already commented in Sec.~\ref{gravitational}.

For our numerical estimates we will fix $\delta$ to its upper bound $\delta=0.1$. The contour plot for fixed values of $\Lambda_S$ (in TeV) is plotted in the left panel of Fig.~\ref{fig5} in the plane $(\omega,K_H)$. The preferred value of $\Lambda_S$ can be obtained from the lower experimental limit on the gluino mass $M_3\gtrsim 1.5$ TeV, which translates into the bound $M_1\gtrsim 250$ GeV for our minimal gauge mediation~\footnote{These bounds could be evaded in non-minimal gauge mediation models, as in models of general gauge mediation~\cite{Meade:2008wd}.}. Then from Eq.~(\ref{Mi}) we can extract the value $\Lambda_S=4\pi M_3/(\sqrt{n_5}\alpha_3(M_3))\gtrsim 230\textrm{ TeV}/\sqrt{n_5}$. 
This value imposes constraints on the hidden sector parameter $K_H$, which should be as large as possible. First of all we see that the value of $K_H$ depends on the value of the unification coupling constant and the number of messengers which are non singlets under the group $G_H$. The gauge coupling unification value for the MSSM is $\alpha_{GUT}^{MSSM}\simeq 1/24$ for a unification scale $M_{GUT}\simeq 2\times 10^{16}$ GeV. The possible presence of $n_{10}$ states in complete representations of $SU(5)$ increases the value of $\alpha_{GUT}$ leaving (at one-loop) the value of the unification scale unmodified, as it was shown in the right panel of Fig.~\ref{fig4}.

After fixing $\delta$ we are then left with the parameters:
\begin{equation}
M, \quad n_5, \quad n_{10}, \quad \textrm{and}  \quad C_5^{G_H}\ .
\end{equation}
In the simplest framework of minimal gauge mediation considered here, and as from the previous considerations, we are left to consider the cases of $n_5=1,2$. Moreover the messenger mass $M$ is constrained to be of order:
\be
M\lesssim \left(0.1/R\right) \sim 10^{10} {\textrm{  GeV}}
\label{cota}
\ee
from the validity of our approximation for $F$, as explained in App.~\ref{apendice}, combined with a conservative value of $\left(1/R_0\right)$ extracted from Fig.~\ref{fig2}. Notice that the requirement $F/M^2<1$ (a condition to avoid tachyons in the messenger sector in gauge mediation) is always satisfied because of the smallness of $\alpha_H$. Clearly, $C^{G_H}_5$ has a strong impact on the value of $K_H$ and the spectrum of the model. This is given by the squared of the charge of the messengers under $U(1)_H$, $Q_H^2$. Typical values could be for example $C^{G_H}_5 =1$ ($Q_5=\pm 1$), in which case the model is more contrived as we shall discuss shortly, or $C^{G_H}_5=4$ ($Q_5=\pm 2$) which gives more room for the other parameters. Examples of models with extra $U(1)$'s and their charges can be found in the literature, see e.g.~Ref.~\cite{Ibanez:1998qp}.

For $C^{G_H}_5=1$, the necessary values of $K_H$ require larger values of $\alpha_{GUT}$. This could be achieved by increasing $n_5$ and/or $n_{10}$ and lowering $M$. However, keeping the Bino as the LSP requires instead lower $n_5$, and higher $M$, values, which creates a tension. Hence, we choose to keep $M$ in the range $10^8 - 10^{10}$ GeV and introduce a number of neutral messengers, $n_{10} > 0$. Examples of valid sets $[M , n_5, n_{10}, \alpha_{GUT}]$  leading to phenomenologically viable spectra are e.g.~$[10^8 \textrm{  GeV}, 1, 1, 1/10]$ or $[10^7 \textrm{  GeV}, 2, 2, 1]$. A larger range for the parameters would of course be allowed for bigger values of $C^{G_H}_5$. The used value of $1/R$ and the corresponding value of $M_s$ according to Eq.~(\ref{Planckrelation}), for the case $\alpha_{GUT}\simeq\mathcal O(1)$, are shown in the right panel of Fig.~\ref{fig5}, from where we can see that $M_s> M_{GUT}$. The value of $M_s$ scales as $\alpha_{GUT}^{2/3}$ so that for $\alpha_{GUT}=1/10$ the value of $M_s$ should be reduced by a factor $\sim 0.22$.

Another way of increasing the value of $K_H$, which we will not pursue along this paper, is to assume that there is an additional number $d_T$ of extra dimensions (where the  group $G_H$ does not propagate) with radii $R_T$ slightly larger than the string length $\ell_s$, in which case the right-hand side of Eq.~(\ref{KH}) is multiplied by $(M_s R_{T})^{d_T/3}$ with the corresponding enhancement factor for $K_H$. This situation can arise for instance in type I strings where the SM is in a $D3$-brane and the large SS dimension inside a $D7$-brane, in which case $d_T=2$. However the presence of large extra dimensions also affects Eq.~(\ref{Planckrelation}) by lowering the value of $M_s$ and we might lose the unification properties at the scale $M_s$.

Finally, we would like to comment about the presence of the secluded $U(1)_H$. At one-loop the messengers could a priori introduce a kinetic mixing between $U(1)_H$ and the hypercharge $U(1)_Y$. This mixing vanishes in the simplest case of messengers in full representations of $SU(5)$ and   common mass as considered above. It is otherwise suppressed by the smallness of the hidden gauge coupling  to be  $\lesssim 10^{-5}$.  A soft mass for $G_H$ charged scalars living on the other end of the large dimension is induced at one-loop by the hidden gaugino mass as in \cite{Antoniadis:1998sd,Delgado:1998qr}. Their vacuum expectation value could then generate a mass for the $U(1)_H$ gauge boson of order $\sim (\textrm{loop factor})^{1/2} g_H /R \sim\mathcal O(10^7 \textrm{ GeV)}$, where $g_H$ is the hidden gauge group coupling. A quantitative discussion of the phenomenological and cosmological implications requires considering explicit models of the hidden sector, a subject which goes beyond the scope of this paper.

\section{$\mu/B_\mu$-terms}
\label{Mu}
It is well known that gauge mediation cannot induce $\mu/B_\mu$ terms, as gauge interactions cannot generate them without direct couplings between the Higgs and messenger sectors~\cite{Giudice:1998bp}. However gravitational interactions could do the job as in the Giudice-Masiero mechanism~\cite{Giudice:1988yz}. In this section we will illustrate this point by assuming that some particular effective operators are generated in the higher dimensional theory. 

The SS breaking can be understood in terms of supersymmetry breaking induced by the $F$ term of the radion~\cite{Antoniadis:1991kh} superfield $\widetilde T$, with bosonic components~\footnote{We will use the dimensionless superfield $T$.} 
\be
\widetilde T/M_s\equiv T=R M_s+\theta^2 m_{3/2} ,\quad F_{\widetilde T}=m_{3/2}M_s
\label{radion}
\ee
where $m_{3/2}$ is the mass of the gravitino zero mode given in Eq.~(\ref{gravmasses}). The radion superfield then induces bulk gaugino $\lambda_H$ masses through the coupling
\be
\mathcal L=\frac{1}{4}\int d^2\theta TW_{H\alpha} W^\alpha_H\ .
\ee
The radion superfield in Eq.~(\ref{radion}) can also induce the $\mu$ and $B_\mu$ parameters required for electroweak breaking through effective operators as
\be
\int d^4\theta \left\{ \left[a^2\,|f(T)|^2+b\, f(T^\dagger)\right]H_1\cdot H_2+h.c.\right\}
\label{effmu}
\ee
where $H_1$ and $H_2$ are the two Higgs doublets of the MSSM, $a$ and $b$ are real parameters and $f(T)$ is a real function~\footnote{One can also introduce in (\ref{effmu}) two different functions $f(T^\dagger)+|g(T)|^2$, but here we trade this freedom for the arbitrariness of the real coefficients $a$ and $b$.}. Using (\ref{effmu}) one can compute the values of the $\mu$ and $B_\mu$ parameters as
\begin{align}
|\mu|&=\,m_{3/2}\left|f^\prime_{T}(RM_s)\right|\left|b+a^2 f(RM_s)  \right|\nonumber\\
B_\mu&=a^2\,m_{3/2}^2 \left|f^\prime_T(RM_s)   \right|^2
\end{align}
where we have to impose the phenomenological condition from electroweak symmetry breaking $|\mu|^2\simeq |B_\mu|$, which requires the relation 
\begin{equation}
\left|b+a^2f(RM_s)\right|\simeq |a|\,.
\label{condicion}
\end{equation}
\begin{figure}[htb]
\begin{center}
 \begin{tabular}{cc}
 \hspace{-0.5cm} \includegraphics[width=8.5cm]{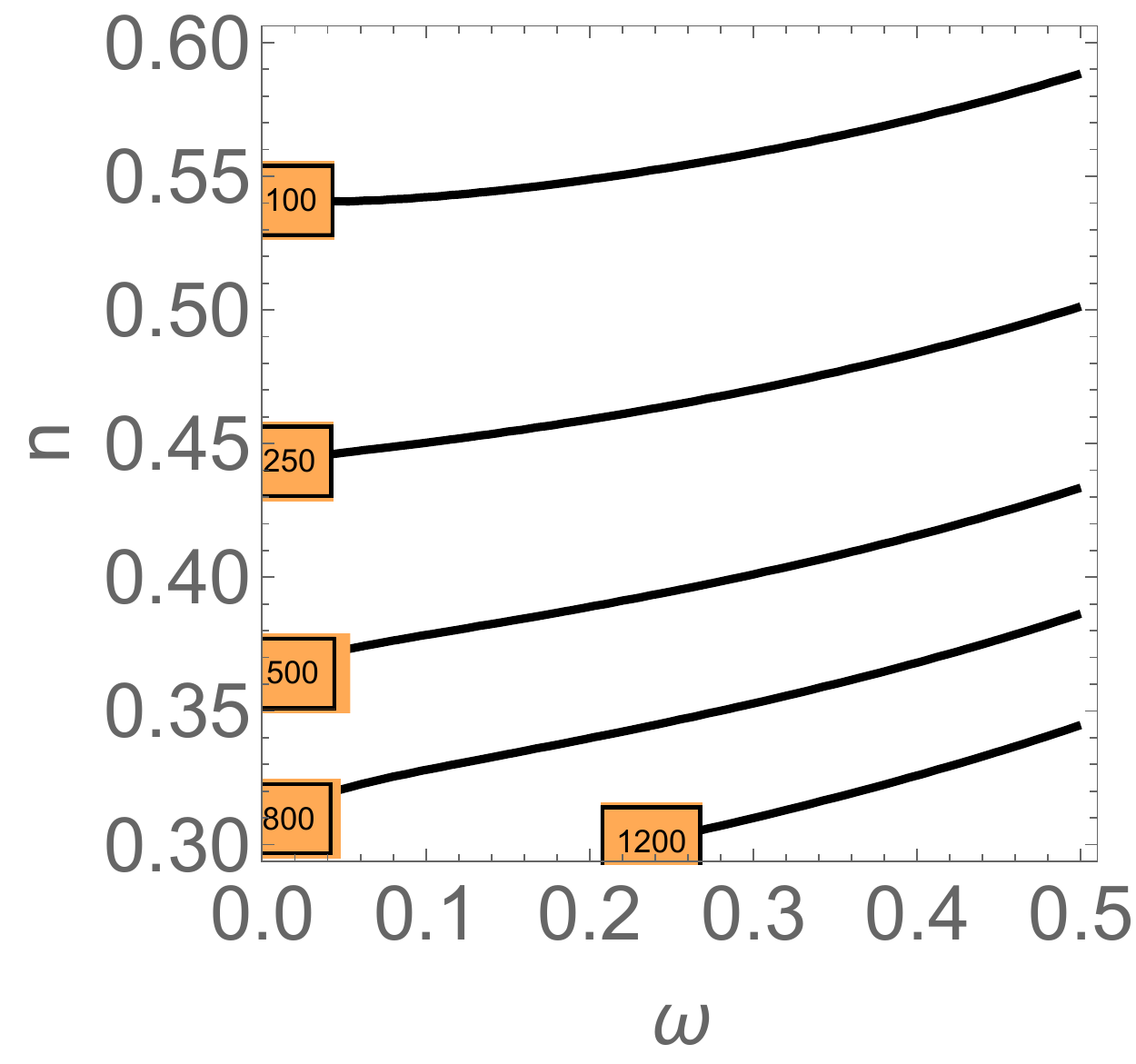}
  \includegraphics[width=9.5cm]{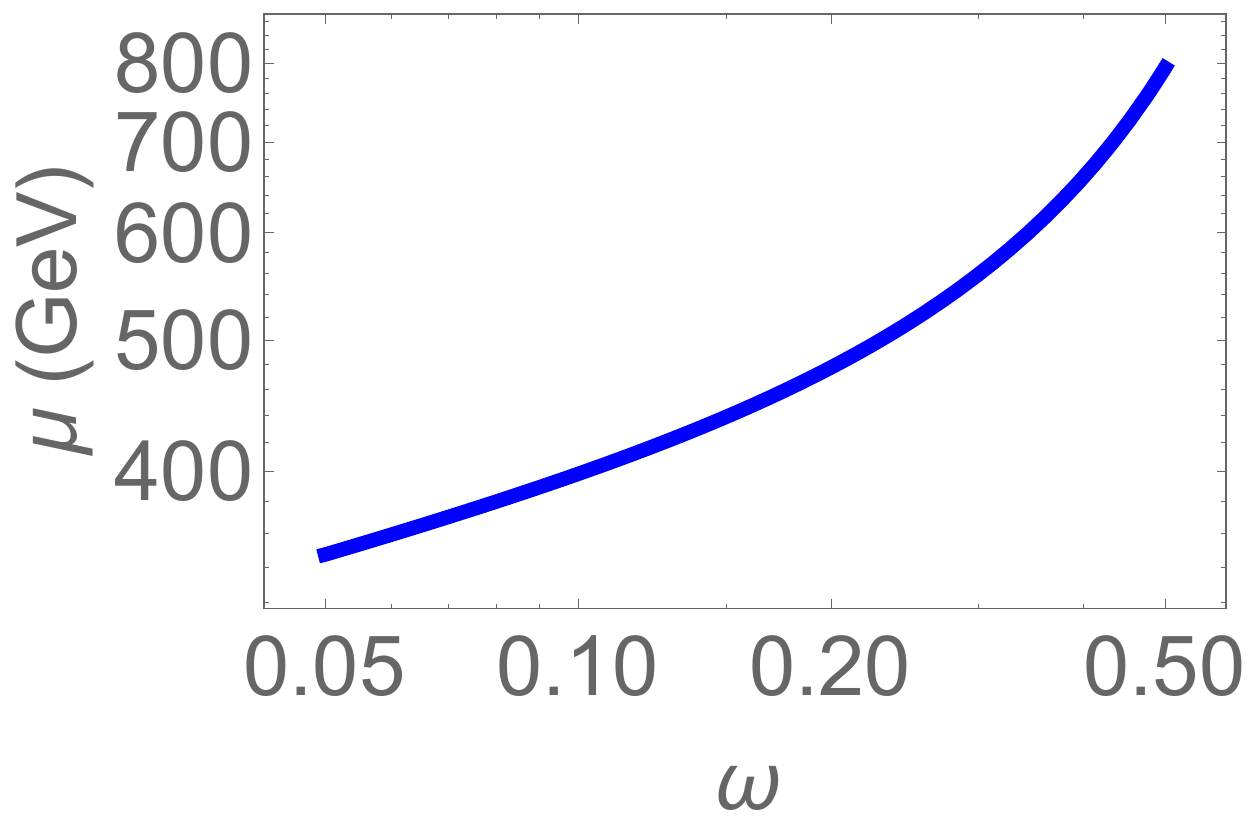}  
\end{tabular}
\end{center}
\caption{\it Left panel: Contour lines of $|\mu|$ in GeV for $a=b=1$ in the plane $(\omega,n)$. Right panel: Plot of $|\mu(\omega)|$ for the case $f(T)=\log(T)$ for $a\simeq b\simeq 1/32\pi^2$.}
\label{fig6}
\end{figure}
For instance by using an asymptotic form (for large $T$) $f(T)\simeq T^{-n}$ and fixing $1/R=1/R(\omega)$ we can determine the value of the $\mu$ parameter for $a\simeq b\simeq 1$ as in the left panel plot of Fig.~\ref{fig6}, where we plot contour lines of constant $|\mu|$ in GeV in the plane $(\omega,n)$.  The case $f(T)\simeq\log (T)$ is also plotted (right panel), for the case $a\simeq b\simeq \frac{1}{32\pi^2}$. As we can see this case only works if both $a$ and $b$ parameters are loop suppressed so that condition (\ref{condicion}) is also satisfied. In all cases the value of $M_s$ from the right panel of Fig.~\ref{fig5}, which corresponds to $\alpha_{GUT}=\mathcal O(1)$, has been used and the electroweak breaking condition $|\mu|^2\simeq B_\mu$ is fulfilled.

\section{Conclusion}

In this work, we presented a novel mechanism of supersymmetry breaking where the SM gaugino, squark and slepton masses arise predominantly from flavor blind gauge mediated interactions, while the gravitino mass is superheavy due to an appropriate sequestering of the supersymmetry breaking sector.  We have presented an example  for how $\mu$ and $B_\mu$ parameters could be generated at the same time by effective supergravity interactions, as in the Giudice-Masiero mechanism. 

Some important questions have been left aside. For instance, the radion stabilization was not discussed here. The potential generated for the modulus $T$ should be such that at its minimum it reproduces the required hierarchy between the compactification and string scales. We are assuming here that the required stabilization mechanism will not perturb the main features of the mechanism presented here. Also, one needs to understand the origin of the effective operators describing the couplings of the radion to matter fields, as assumed in Sec.~4. We believe that such issues should be addressed in a more fundamental theory. The proposed mechanism should be in principle realized in string theory but all calculations have been done in the context of the effective field theory by summing over the contribution of the tower of KK modes. This can be tested by an explicit string implementation and model building which is left for future research as a very interesting open problem.

\appendix

\section{Calculation of $\Pi(q^2)$}
\label{apendice}
In Sec.~\ref{gauge} we have computed the seed of supersymmetry breaking for the calculation of gauge mediation as $F=\Pi(0)$ for zero external momentum. In this appendix we will compute $\Pi(q^2)$ for arbitrary external momentum $q$. To simplify the notation we will use inn this section units in which $1/R\equiv 1$ so all masses $m$ are scaled as $mR$. We will also consider the case where $M\ll 1$ as in Sec.~\ref{gauge}. The integral (\ref{integral}) is then generalized to 
\be
F(q)=\frac{\alpha_H}{\pi}C_H^\phi I(q)
\ee
where
\be
I(q)=\int_0^\infty dp^2 p^2 \sum_n\left[\frac{M_n(\omega)}{p^2+M_n^2(\omega)}\frac{M}{(p+q)^2+M^2} -\frac{M_n(0)}{p^2+M_n^2(0)} \frac{M}{(p+q)^2+M^2}\right] 
\ee
and rotation to Euclidean momenta is performed. We now introduce integration over $\int_0^1dx$ and the change of integration momentum $p\to p-q(1-x)$ and $p^2\to x\,p^2$
\begin{equation}
I(q)=\int_0^1dx\int_0^\infty dp^2 p^2 \sum_n\left[\frac{M_n(\omega)M}{\left[M_n^2(\omega)+A\right]^2}-\frac{M_n(0)M}{\left[M_n^2(0)+A\right]^2}
\right]
\end{equation}
where
\be
A=p^2+B,\quad B=q^2(1-x)+M^2(1-x)/x
\ee

\begin{figure}[htb]
 \begin{tabular}{cc}
 \includegraphics[width=7.5cm,height=6.5cm]{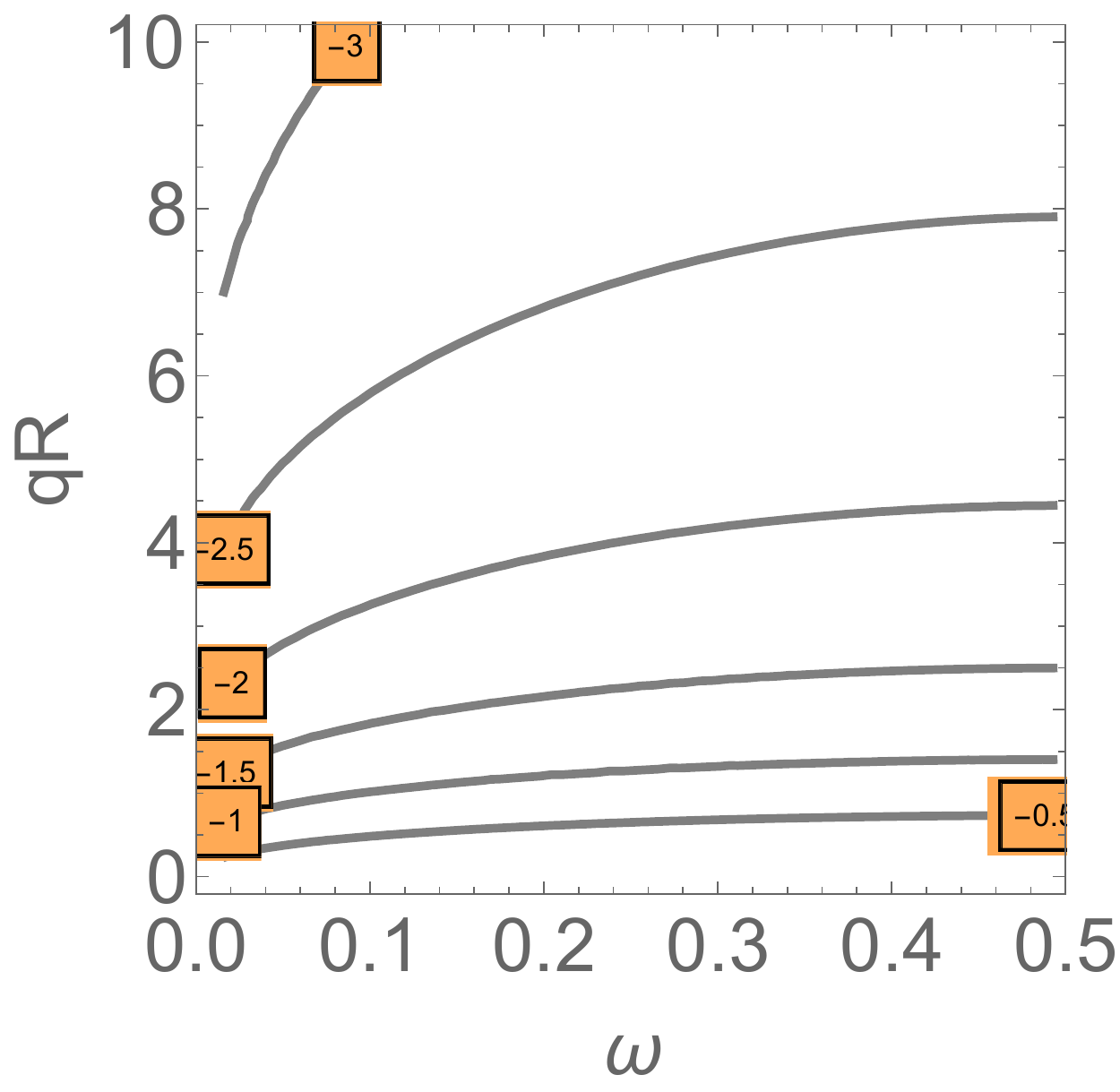}  
  \includegraphics[width=11.cm,height=6cm
  ]{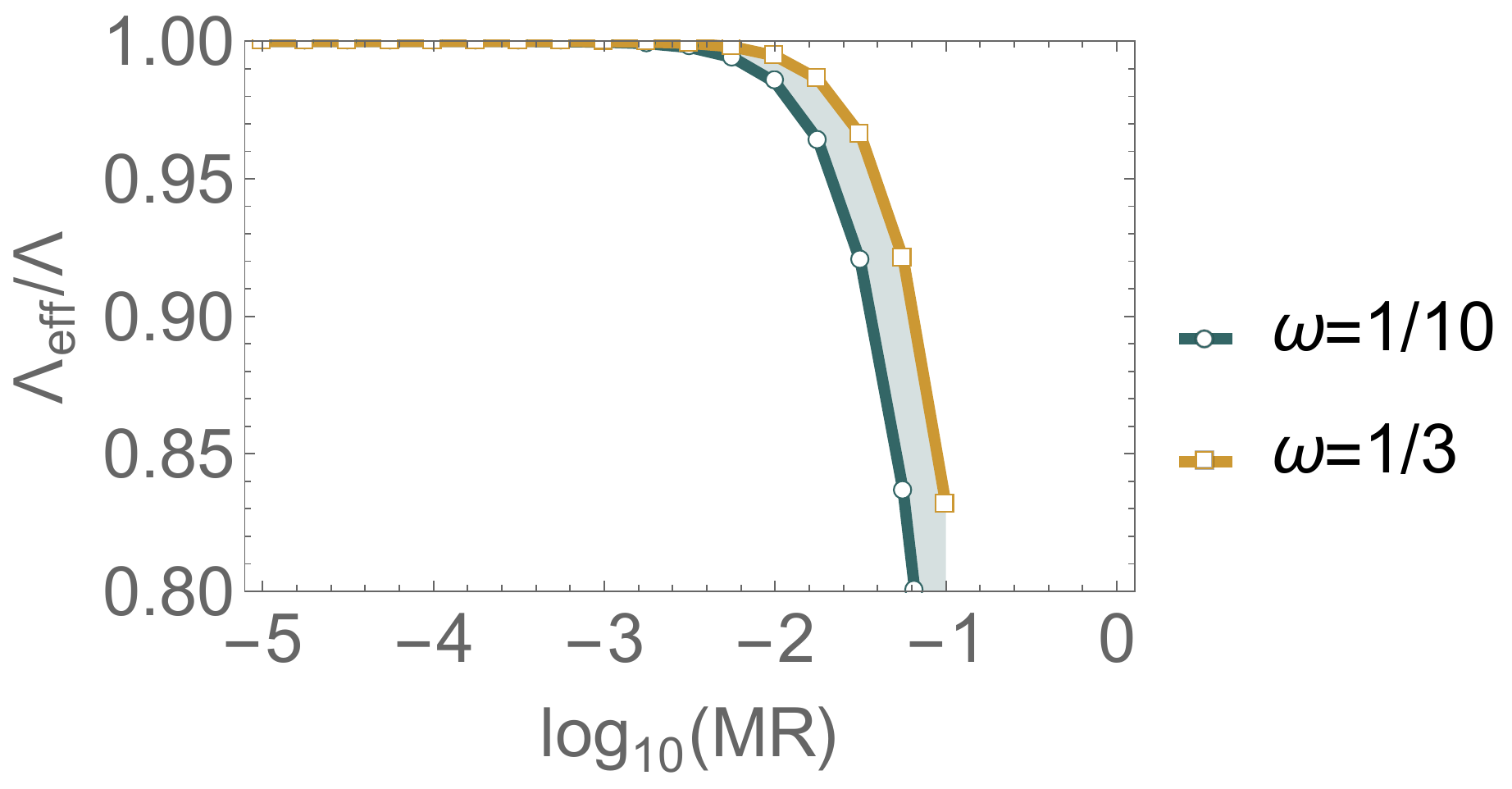}  
\end{tabular}
\caption{\it Left panel: Contour lines of $\log_{10}[F(q)/F(0)]$. Right panel: Plot of $\Lambda_{eff}/\Lambda$ for different values of $\omega$.}
\label{fig7}
\end{figure}

After performing the summation over $n$ and the integration over $p$ one easily obtains
\be
I(q)=i\,\int_0^1 dx \left\{\sqrt{B}\log\left[ 1-e^{2\pi(\sqrt{B}-i\omega)}\right]+\frac{1}{2\pi}
Li_2\left( e^{2\pi(\sqrt{B}-i\omega)}\right)+h.c.\right\}
\ee
In particular, for $M\ll 1$, $\sqrt{B}=q\sqrt{1-x}$, $I(0)=g_0(\omega)/4$ and we recover (as we should) the corresponding case studied in Sec.~\ref{gauge} [see Eq.~(\ref{F})]. Moreover the function $I(q)$ is dominated by its value at the $q=0$ region as we can see in the left panel of Fig.~\ref{fig7} where we plot contour lines of $\log_{10}[F(q)/F(0)]$ in the plane $(\omega,q)$. We can see that, for any value of $\omega$, the function $F(q)$ is dominated by its value at $q=0$.

In gauge mediation we define the supersymmetry breaking parameter $\Lambda\propto F/M$ which seeds the gaugino and scalar masses. We can compare $\Lambda$ with the value $\Lambda_{eff}$ induced after integration of $F(q)$ over the one-loop (for gaugino masses) or two-loop (for scalar masses) diagrams with internal momentum $q$. In view of the result of the left panel of Fig.~\ref{fig7} we expect that $\Lambda_{eff}/\Lambda\simeq 1$ for $MR\ll 1$ as, in the integration over $q$, the momentum $q$ is rescaled as $q\to qM$. This result is quantified in the right panel of Fig.~\ref{fig7} where we have computed the value of the parameter $\Lambda_{eff}/\Lambda$ when $\Lambda_{eff}$ is contributing to the one-loop gaugino masses. We can see that $\Lambda_{eff}/\Lambda\simeq\mathcal O(1)$ is fulfilled for $MR\lesssim 0.1$. Then as we got from the right panel of Fig.~\ref{fig5} that $1/R\lesssim 10^{11}$ GeV, the previous condition translates into the mild condition $M\lesssim 10^{10}$ GeV, a region where the approximation done in Sec.~\ref{gauge} is  fully justified.

\section*{Acknowledgments}

\noindent 
 K.~B.~acknowledges support from the ERC advanced grant ERC Higgs@LHC, the ANR contract HIGGSAUTOMATOR, the Institut Lagrange de Paris (ILP) and the ICTP-SAIFR for hospitality during part of this project. The work of M.Q.~is
partly supported by the Spanish Consolider-Ingenio 2010 Programme CPAN
(CSD2007-00042), by MINECO under Grants CICYT-FEDER-FPA2011-25948 and CICYT-FEDER-FPA2014-55613-P, by the Severo Ochoa
Excellence Program of MINECO under Grant SO-2012-0234, by
Secretaria d'Universitats i Recerca del Departament d'Economia i
Coneixement de la Generalitat de Catalunya under Grant 2014 SGR 1450 and by CNPq PVE fellowship project under Grant number 405559/2013-5. M.Q. acknowledges the LPTHE-ILP, where part of this work has been done, for partial financial support.

\end{document}